# Argumentation Mining: Exploiting Multiple Sources and Background Knowledge

Anastasios Lytos[1], Thomas Lagkas[2], Panagiotis Sarigiannidis[3], and Kalina Bontcheva[4]

[1]Department of Computer Science, South-East European Research Centre, The University of Sheffield, Thessaloniki, Greece
[2]Computer Science Department, The University of Sheffield International Faculty, CITY College, Thessaloniki, Greece
[3]Department of Informatics and Telecommunications Engineering, University of Western Macedonia, Kozani, Greece
[4]Department of Computer Science, The University of Sheffield, Sheffield, UK

**Abstract.** The field of Argumentation Mining has arisen from the need of determining the underlying causes from an expressed opinion and the urgency to develop the established fields of Opinion Mining and Sentiment Analysis. The recent progress in the wider field of Artificial Intelligence in combination with the available data through Social Web has create great potential for every sub-field of Natural Language Process including Argumentation Mining.

## 1 Introduction

Argumentation Mining (AM) is a multi-disciplinary field that first introduced in 2009 from Palau and Moens [20] and gained the interest of the scientific community because of the progress in many fields in Artificial Intelligence (AI) mostly due to the development of Machine Learning (ML) techniques, algorithms and platforms. Another reason for the development of the field is the explosion in the use of Social Media and other interactive capabilities such as comments sections, on-line product reviews and personal blogs.

In human reasoning interpreting an argument is a natural process that is realized automatically by analyzing many different aspects regarding the discussing topic. However, modeling the argumentation process for the purpose of Automatic Argumentation Mining is a challenging task with doubts that is even feasible.

The main reason behind the effortless and instant grasp of the underlying argument behind from an expressed opinion is the capability of the human being to perceive the context of the expressed opinion by combining multiple sources of information. The modeling and the exploitation of the background knowledge is a difficult task especially if we consider hardware limitations both in memory and speed.

The problems related to AM become more severe when web-generated data are used as the use of slang is quite often, abbreviations are frequent and there are fallacies in the reasoning process. The aforementioned problems are the main reason behind the exploitation of domains that have structured and standardized reasoning for the tasks of AM such as law [28, 33] and scientific text [5, 12, 15].

The exploitation of data generated from Web is a more challenging task comparing to AM in structured data as the data are unstructured, the expressed opinions are short and quite often they don't include an argument. Opinion Mining and Sentiment Analysis are thriving in the era of Social Web, however AM seems not to be able to exploit the capabilities from the volume and variety of data that are offered.

In order to take advantage of the new capabilities offered from the Web in the field of AM we propose alternative paths from the implementation of the existing Argumentation theories. We propose the development of two new schemes for the exploitation of Web-generated data and small texts in general, relying on two basic principles 1) the constant production of background knowledge 2) the combination of multiple sources and the evaluation of them. In this paper we propose two abstract frameworks that are characterized by the capabilities of modification and extensibility.

## 2  Related Work

AM as a research field is quite recent but the study and the analysis on arguments in the speech are held from 4 century B.C. [2] changing forms and objectives through years. In this section we will present the Argumentation Models which have proposed at the end of the 20th century and the first attempts of AI implementing Argumentation Models before proceeding to the task of Automatic AM.

The illustration of argument parts with the use of interconnected nodes is a common technique nowadays and first introduced by Beardsley in [4]. In Beardsley's model there are defined three basic categories of arguments: 1) Convergent Argument 2) Divergent Argument and 3) Serial Argument. Beardsley's theory laid the foundation of many recent Argumentation Schemes, however it has the weakness of not defining the relationships between the nodes.

A more detailed scheme is introduced from Toulmin [30] which stands out because of the precise description of the argument entities. In Toulmin's model six functional roles were suggested, datum, warrant, backing, qualifiers and rebuttal. These roles provide a quite detailed view of the expressed arguments, showing the completeness or not of an argument. In figure 1 an example of Toulmin's theory is depicted where datum is the first part of the argument in the left, the warrant part is expressed through the conjuction since, the backing supports the warrant and both of them lead to the qualifier. The rebuttal part is optional and provides additional support to qualifier.

A different approach is followed in Mann and Thompson [18] aiming at the organization of the text into different regions. The proposed architecture is characterized as an open scheme with few established rules and is offered for extensions and modifications. The basic concepts of the scheme is a central part under the name *nucleus* which is framed with a number of *satellites*. The distinction of nucleus-satellite is applied recursively until every part of the argument is associated with another one. The relations that have created are depicted either in tree-structure format either in XML format [27].

The first attempts on modeling arguments with AI techniques and models took place in the 1990's with the pioneer research of Pollock [24] which describes the connection between philosophy non-monotonic reasoning in AI. These first attempts were focused in fields where communication takes place according a series of well-established rules of communication such as law, rhetorics and scientific text.

The connection between argumentation and Logic Programming was researched in Dung [9] and in Krause et al. [16]. In both attempts the approach that was followed is the establishment of a series of rules, definitions and propositions aiming at the acceptability and the integrity of the reasoning process. A similar approach was also followed from Parsons and Jennings [22] without focusing on the optimum solution but on an acceptable compromise.

The supremacy of tailored arguments in an advising system was researched in Carenini and Moore [8] based on the generation system they have established earlier [7]. A similar problem, the solution of possible conflicts in a dialectical argumentation schemes, is researched on the work of Grasso et al. [11] with the implementation of the theory that was developed from Perelman and OlbrechtsTyteca [23].

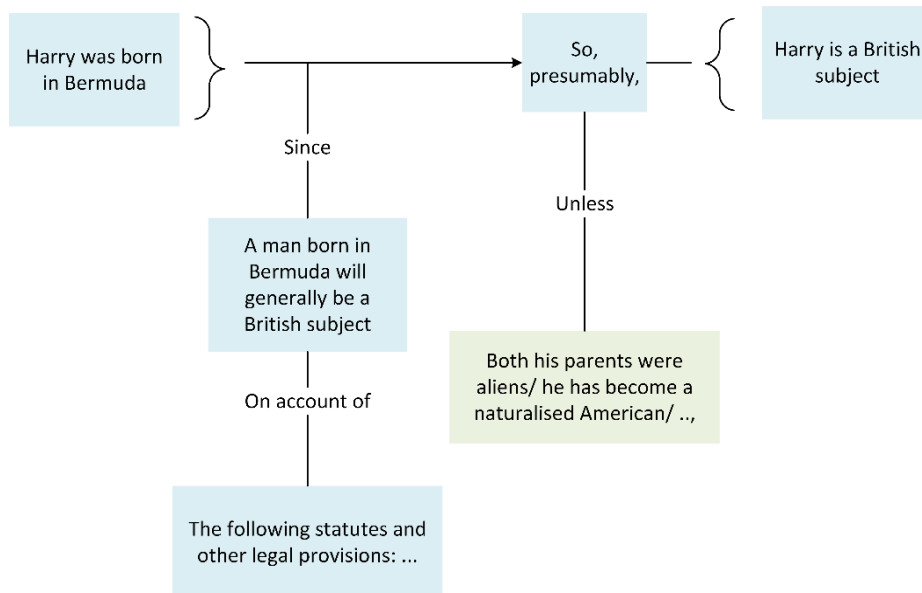

**Fig. 1.** Toulmin's Scheme Example [30]

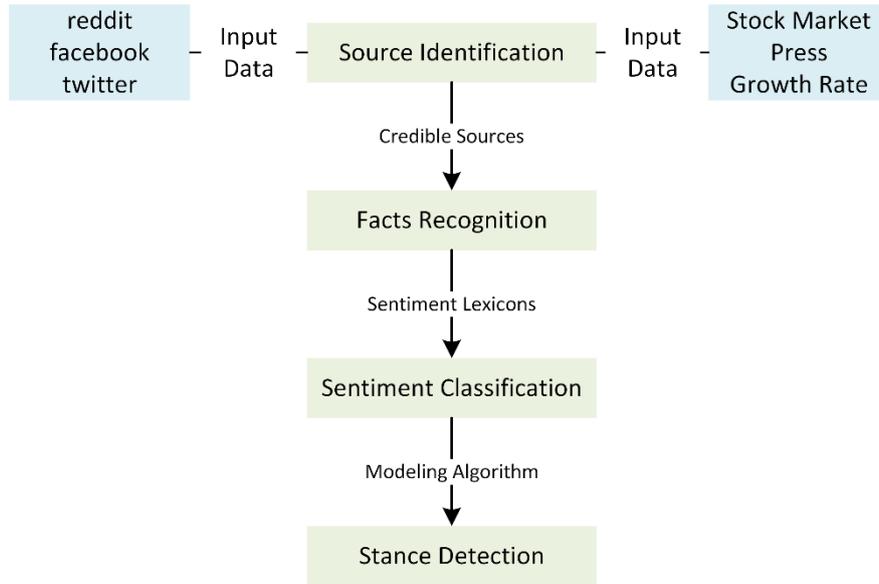

**Fig. 2.** Proposed Architecture for Combination of Multiple Sources

## 3   Argumentation Mining and Automatic Classification

The term of AM refers to many different interrelated tasks that can be research either independently either in the wider context of extracting arguments from text. The automatic classification of arguments is the last step of the AM pipeline and usually is the outcome of the previous steps applied on a specific task on a specific field. There is recent work which exploits Internet sources like twitter [1, 10, 19, 29], reddit [13] and other forums [6, 17, 21, 31] but none of the previous work do not combine the Internet unstructured data with objective and measurable metrics.

### 3.1   Combining Multiple Sources

In this subsection we propose a scheme which combines multiple sources in order to assess the reliability of them. The proposed pipeline exceeds the established boundaries of AM as is it can be applied only to commercial problems revealing real processing mechanisms of the human brain. Eventually the proposed pipeline examines the question of quantity vs quality as the core of the research is to find what affects more the general opinion, the plethora of anonymous opinions or thoughtful opinions and objective metrics.

In figure 2 the proposed pipeline is presented in an abstract from which can be modified and extended. In order to illustrate better the proposed architecture we will provide step-by-step directions of every step. Assuming a controversial topic that has intrigued

69

both the Social Media platforms and traditional media like the recent privacy scandal of facebook1.

The first step of the proposed pipeline is the collection of data through the conversation on Social Media, articles of recognized media in their digital format and the monitoring of the stock market for the specific firm and its competitors. Provided that the collection of the data has completed the second step is the *Source Identification* of each input data followed by the stage of *Facts Recognition*. These two steps are crucial in our analysis as through research the appropriate weight will be assigned and form the final pipeline. The next step is the *Sentiment Classification (SC)* where different sentiment lexicons evaluate the collected data, SC is a well-known challenge for the scientific community and different lexicons have created [3, 14, 25, 26, 32]. The ultimate step of the proposed pipeline is the *Stance Detection* where the outcome of the architecture is finally formed and the degree of success of the modeling process is evaluated.

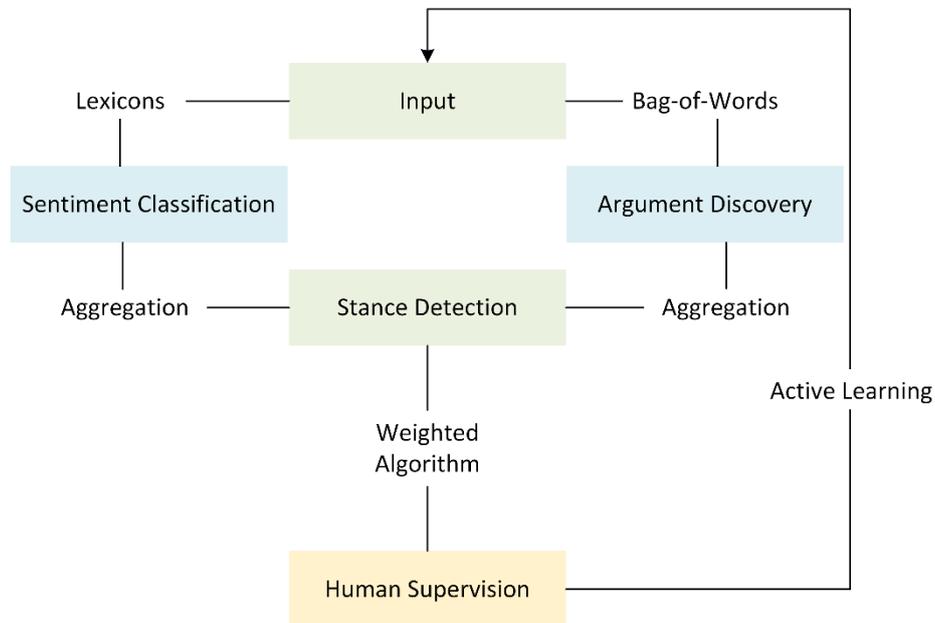

**Fig. 3.** Proposed Architecture for Generating Background Knowledge

### 3.2    Exploiting and Adapting Background Knowledge

Apart from the combination of multiple sources the second challenge we try to answer is the constant generation and adaptation of the background knowledge of the system. In the process of the human reasoning the interpretation of the natural language is an

---
[1] https://www.theverge.com/2018/4/9/17214814/facebook-data-notification-cambridge-analytica

70

instant operation that combines the linguistic, lexical and sentiment characteristics of the language with the background knowledge that was obtained previously from multiple sources. The most remarkable process of the human brain is the adaptation of the background knowledge depending from the subject of the discussion, the reliability of the source and the completeness of the argument itself.

In the proposed architecture, depicted in fig. 3, a constant feedback of the AM pipeline is presented for the dynamic adaptation of it. The feedback loop is based on the interaction of the human factor, which evaluates the model and proceeds in the redesigning of the system if necessary. In this way the pipeline is evaluated regularly and remains up-to-date with possible changes in the complicated environment it interacts.

For the clarification of the proposed scheme we will proceed in a step-by-step example. We have the input collected for a specific topic (e.g. facebook privacy scandal) followed by two processes, the Argument Discovery and the Sentiment Classification. In the process of Argument Discovery the effort is focused in the existence or not of an argument in the text, whereas the Sentiment Classification provide the information for sentiment tone of the text. These two processes lead to the step of the Stance Detection where the arguments in a specific topic are summarized and eventually in the Modeling step their significance and influenced are determined. The last step of the suggested model is the evaluation of it from a human perspective which could find flaws and defects in the system and could proposed modifications.

The ultimate target of AM is the automatic interpretation, evaluation and classification of arguments. However, every suggested pipeline requires the human supervision for evaluation of the proposed model, thus neither our model could avoid the human interaction.

## 4   Conclusions and Future Work

In this short paper we provide a solid background in the area of AM covering different aspects and presenting the evolution of the Argumentation problem through time and different scientific views. We stressed the significance of the field for the better understanding of the natural language and identified the prosperous environment for the growth of the field in the era of Social Web due to the huge amount of web-generated data.

In the third section of the paper we propose two schemes in an effort to interpret and exploit the constantly changing environment of the Web providing new ideas and schemes for the task of AM. Concerning the proposed pipeline for the exploitation of different sources of input data, the pipeline is open to modification and able to cover different problems such as the location of underlying relations between crowd-sourcing and stock market or the effect of different sources of news. The second scheme stress the need for more adaptable models in the field of AM which can provide feedback from the results instead of stiff architectures that cannot adapt to new requirements and needs.

We strongly believe that the field of AM could flourish in the era of Social Web provided that the scientific community can adapt to the new conditions and exploit the new opportunities. The proposed AM schemes should be flexible considering arguments with flaws or weakly expressed and presenting solutions for the industry such as precise recommendation system or advertising impact in Social Media. Apart from industry applications another concern AM has to face technical challenges like Big Data, Deep Learning and Unsupervised Machine Learning Techniques, technologies that could boost the field of AM and the wider area of Natural Language Process.

## References

1. Addawood, A. A. and Bashir, M. N. (2016). What is Your Evidence? A Study of Controversial Topics on Social Media. In Proceedings of the 3rd Workshop on Argument Mining, pages 1–11, Berlin, Germany.
2. Aristotle and Kennedy, G. A. (2006). On Rhetoric: A Theory of Civic Discourse.
3. Baccianella, S., Esuli, A., and Sebastiani, F. (2010). SentiWordNet 3.0: An Enhanced Lexical Resource for Sentiment Analysis and Opinion Mining. In Nicoletta Calzolari (Conference Chair) and Khalid Choukri and Bente Maegaard and Joseph Mariani and Jan Odijk and Stelios Piperidis and Mike Rosner and Daniel Tapias, editor, Proceedings of the Seventh International Conference on Language Resources and Evaluation (LREC'10), Valletta, Malta.
4. Beardsley, M. C. (1950). Practical Logic. The Philosophical Quarterly.
5. Blake, C. (2010). Beyond genes, proteins, and abstracts: Identifying scientific claims from full-text biomedical articles. Journal of Biomedical Informatics, 43(2):173–189.
6. Boltuzic, F. and Snajder, J. (2014). Back up your Stance: Recognizing Ar-ˇ guments in Online Discussions. In Proceedings of the First Workshop on Argumentation Mining, pages 49–58, Stroudsburg, PA, USA. Association for Computational Linguistics.
7. Carenini, G. (2000). Generating and Evaluating Evaluative Arguments. PhD thesis, University of Pittsburgh.
8. Carenini, G. and Moore, J. D. (2001). An empirical study of the influence of user tailoring on evaluative argument effectiveness. In IJCAI'01 Proceedings of the 17th international joint conference on Artificial intelligence - Volume 2, pages 1307–1312, Seattle, WA, USA.
9. Dung, P. M. (1995). On the acceptability of arguments and its fundamental role in nonmonotonic reasoning, logic programming and n-person games. Artificial Intelligence, 77(2):321–357.
10. Dusmanu, M., Cabrio, E., and Villata, S. (2017). Argument Mining on Twitter: Arguments, Facts and Sources. In Proceedings of the 2017 Conference on Empirical Methods in Natural Language Processing, pages 2317–2322, Copenhagen, Denmark.
11. Grasso, F., Cawsey, A., and Jones, R. (2000). Dialectical argumentation to solve conflicts in advice giving: a case study in the promotion of healthy nutrition. International Journal of Human-Computer Studies, 53(6):1077– 1115.

12. Guo, Y., Korhonen, A., and Lattice, T. P. (2011). A Weakly-supervised Approach to Argumentative Zoning of Scientific Documents. In Proceedings of the 2011 Conference on Empirical Methods in Natural Language Processing, pages 273–283, Edinburgh, Scotland.
13. Hidey, C., Musi, E., Hwang, A., Muresan, S., and McKeown, K. (2017). Analyzing the Semantic Types of Claims and Premises in an Online Persuasive Forum. In Proceedings of the 4th Workshop on Argument Mining, pages 11–21, Copenhagen, Denmark. Association for Computational Linguistics.
14. Hu, M. and Liu, B. (2004). Mining and summarizing customer reviews. In Proceedings of the 2004 ACM SIGKDD international conference on Knowledge discovery and data mining - KDD '04, page 168, New York, New York, USA. ACM Press.
15. Kirschner, C., Eckle-Kohler, J., and Gurevych, I. (2015). Linking the Thoughts: Analysis of Argumentation Structures in Scientific Publications. In Proceedings of the 2nd Workshop on Argumentation Mining, pages 1–11, Denver, Colorado.
16. Krause, P., Ambler, S., Elvang-Goransson, M., and Fox, J. (1995). A Logic of Argumentation for Reasoning under Uncertainty. Computational Intelligence, 11(1):113–131.
17. Liebeck, M., Esau, K., and Conrad, S. (2016). What to Do with an Airport? Mining Arguments in the German Online Participation Project Tempelhofer Feld. In Proceedings of the Third Workshop on Argument Mining (ArgMining2016), pages 144–153, Stroudsburg, PA, USA. Association for Computational Linguistics.
18. Mann, W. C. and Thompson, S. A. (1987). Rhetorical Structure Theory: Description and Construction of Text Structures. In Natural Language Generation, pages 85–95. Springer Netherlands, Dordrecht.
19. Mohammad, S. M., Sobhani, P., and Kiritchenko, S. (2017). Stance and Sentiment in Tweets. ACM Transactions on Internet Technology, 17(3):1–23.
20. Palau, R. M. and Moens, M.-F. (2009). Argumentation mining. In Proceedings of the 12th International Conference on Artificial Intelligence and Law ICAIL '09, page 98, New York, New York, USA. ACM Press.
21. Park, J. and Cardie, C. (2014). Identifying Appropriate Support for Propositions in Online User Comments. pages 29–38.
22. Parsons, S. D. and Jennings, N. R. (1996). Neogotiation Through Argumentation - A Preliminary Report. In 2nd Int. Conf. on Multi-Agent Systems, pages 267–274, Japan.
23. Perelman, C. and Olbrechts-Tyteca, L. (1969). The new rhetoric: a treatise on argumentation. University of Notre Dame Press.
24. Pollock, J. L. (1987). Defeasible reasoning. Cognitive Science, 11(4):481– 518.
25. Potts, C. (2010). On the negativity of negation. In Semantics and Linguistic Theory, volume 20, page 636.
26. Potts, C. (2011). Developing adjective scales from user-supplied textual metadata. In NSF Workshop on Restructuring Adjectives in WordNet, Arlington, VA, USA.
27. Reitter, D. and Stede, M. (2003). Step By Step: Underspecified Markup in Incremental Rhetorical Analysis. In Proceedings of the 4th International Workshop on Linguistically Interpreted Corpora (LINC-03), pages 77–84.

28. Savelka, J. and Ashley, K. D. (2016). Extracting Case Law Sentences for Argumentation about the Meaning of Statutory Terms. In Proceedings of the 3rd Workshop on Argument Mining, pages 50–59.
29. Schneider Nathan, O'Connor Brendan, Saphra Naomi, Bamman David, Faruqui Manaal, Smith A. Noah, Dyer Chris, and Baldridge Jason (2013). A Framework for (Under)specifying Dependency Syntax without Overloading Annotators - Semantic Scholar.
30. Toulmin, S. E. (1958). The Uses of Argument.
31. Wei, Z., Xia, Y., Li, C., Liu, Y., Stallbohm, Z., Li, Y., and Jin, Y. (2016). A Preliminary Study of Disputation Behavior in Online Debating Forum. In Proceedings of the 3rd Workshop on Argument Mining, pages 166–171, Berlin, Germany.
32. Wiebe, J. and Mihalcea, R. (2006). Word sense and subjectivity. In Proceedings of the 21st International Conference on Computational Linguistics and the 44th annual meeting of the ACL - ACL '06, pages 1065–1072, Morristown, NJ, USA. Association for Computational Linguistics.
33. Wyner, A., Mochales-Palau, R., Moens, M.-F., and Milward, D. (2010). Approaches to Text Mining Arguments from Legal Cases. In Semantic Processing of Legal Texts, pages 60–79. Springer, Berlin, Heidelberg.